\documentclass{article}

\usepackage[preprint,nonatbib]{neurips_2026}
\makeatletter
\renewcommand{\@noticestring}{}
\makeatother
\usepackage[utf8]{inputenc}
\usepackage[T1]{fontenc}
\usepackage{hyperref}
\usepackage{url}
\usepackage{microtype}

\usepackage{amsmath,amssymb,amsthm,mathtools}
\usepackage{algorithm}
\usepackage{algpseudocode}
\usepackage{graphicx}
\usepackage{subcaption}
\usepackage{wrapfig}
\usepackage{float}
\usepackage{booktabs}
\usepackage{multirow}
\usepackage{longtable}
\usepackage{xcolor}
\usepackage{listings}
\usepackage{tikz}
\usetikzlibrary{arrows.meta,positioning,fit}

\definecolor{promptbg}{RGB}{247,249,252}
\definecolor{promptframe}{RGB}{209,217,230}
\definecolor{prompttext}{RGB}{38,44,59}
\definecolor{promptaccent}{RGB}{17,97,167}
\lstdefinestyle{promptstyle}{
  basicstyle=\ttfamily\scriptsize\color{prompttext},
  backgroundcolor=\color{promptbg},
  frame=single,
  rulecolor=\color{promptframe},
  frameround=tttt,
  breaklines=true,
  showstringspaces=false,
  columns=fullflexible,
  keepspaces=true,
  xleftmargin=1ex,
  xrightmargin=1ex,
  aboveskip=0.6em,
  belowskip=0.2em,
  emph={Objective,Candidate,Metadata,Input,Context,Baseline,Analysis,Optimization,Prescription,Kernel,Synthesis,Target,Policy,Constraints,Recent,Benchmark,Metrics,Profiling,Summary,Bottleneck,Diagnosis,Debug,Logs,Output,Requirements,Current,Source},
  emphstyle=\bfseries\color{promptaccent}
}

\usepackage[nameinlink,noabbrev]{cleveref}

\theoremstyle{definition}

\title{KernelBrain: Coarse-to-Fine, Budget-Aware Search for Agentic GPU Kernel Optimization}
\author{%
Shuai Che\\
Microsoft\\
\And
Gang Peng\\
Microsoft\\
}

\begin{document}

\maketitle

\begin{abstract}
Automating GPU kernel optimization remains difficult in practice: generated
variants can violate correctness constraints, runtime measurements are noisy,
and search often stalls early. We present a practical optimization agent that
combines LLM-guided mutation, adaptive resource allocation, policy-gated
evaluation, and profiler-informed diagnosis. The system screens many candidates
with low-cost evaluation and allocates higher-fidelity budget only to promising
survivors to optimize and evolve GPU kernels.
On important Triton kernel generation tasks, this design improves both kernel quality
and search efficiency, reaching 0.88x--6.72x speedup over PyTorch and
up to 1.4x speedup over the state-of-the-art kernel agent, with up to 48\%
lower optimization time.
\end{abstract}

\section{Introduction}
Efficient large-scale model deployment relies heavily on
high-performance GPU kernels, exemplified by specialized attention mechanism and 
serving libraries such as TensorRT-LLM, vLLM, and SGLang \cite{trtllm,vllm,sglang,flashattention,flashinfer}. However, the manual
development of these kernels has become a significant bottleneck due to the
expansive design space of algorithms, parallelism, memory layouts, and
synchronization primitives. This complexity is compounded by the rapid pace of
hardware innovation, where architectural shifts frequently invalidate established
optimization strategies. While automated evolutionary frameworks
\cite{openevolve} have emerged to bridge this gap, they often lack the
capabilities necessary for deep structural refactoring,
as these methods typically prioritize immediate, incremental performance gains.

Recent LLM-based systems can generate strong kernel variants, yet production
deployment still faces three recurring issues: correctness regressions,
unstable ranking under noisy measurements, and search stagnation. These
problems are amplified by the ``non-monotonic'' nature of the optimization
landscape, where small code changes can shift bottlenecks across compute,
memory, and scheduling---producing large, hardware-dependent performance
swings---and intermediate transformations may be neutral or even temporarily
detrimental to performance. To address these concerns, diagnosis-driven, multi-agent approaches with
hardware-in-the-loop feedback and memory-based reflection can achieve effective
kernel generation \cite{kernelagent_blog, ksearch}, but they often consume substantial resources
on profiling and exploration.

Our goal is therefore not only to find a fast kernel once, but to maintain a
stable, budget-efficient optimization-evolving loop under noise, interruptions, and
heterogeneous workloads. Our main contributions are:
\begin{enumerate}
  \item \textbf{Asymmetric resource allocation.} Kernel
    performance can vary by orders of magnitude between the best and worst
    implementations. We allocate minimal resources to rapidly scan and prune
    underperforming candidates---gated by explicit policy and correctness
    checks with stability adjudication---and concentrate the majority of the
    evaluation budget on the most promising data points through a multi-fidelity
    regime;
  \item \textbf{Expert-guided coarse-to-fine search with hardware profiling.}
    Rather than relying solely on blind LLM-based evolutionary mutation, we employ
    expert-driven hints to cover the major optimization spaces broadly in early
    rounds and progressively narrow the search toward the best-performing
    regions, informed by GPU kernel profiling to provide traceable,
    bottleneck-aware mutation guidance; and
  \item \textbf{State-of-the-art Triton kernel performance.} We evaluate
    the system on Triton-based workloads and demonstrate competitive results,
    achieving $0.88\times$--$6.72\times$ speedup over PyTorch and
    up to 1.4x speedup over KernelAgent, with up to 48\%
    geometric mean) across
    six important DNN benchmarks (Section~\ref{sec:eval}).
\end{enumerate}

\section{Background and Related Work}
Kernel optimization has evolved from hand-tuned templates to learned
autotuning and, more recently, LLM-assisted program search. Classical
autotuners are effective when schedule parameters are fixed in advance, but
they are less effective when improvements require structural code edits.
LLM-based approaches expand the search space, but they also increase the need
for disciplined orchestration: without strong gating and measurement controls,
systems can overfit noise or drift toward low-performing programs.

The broader background of auto-optimization falls into three lines. Compiler and autotuning systems
such as TVM, AutoTVM, Ansor, and Halide learn or search over parameterized
schedule spaces, while Triton provides a practical kernel language that exposes
blocked GPU programs to both human experts and automated systems
\cite{tvm,autotvm,ansor,halide_auto,triton}. Workload-specific systems (e.g., LLM) such as
FlashAttention and FlashInfer show that large gains often come from
hardware-conscious design rather than only parameter tuning
\cite{flashattention,flashinfer}. More generally, AlphaTensor,
AlphaEvolve, OpenEvolve, motivate viewing optimization as evaluator-
driven search \cite{alphatensor,alphaevolve_agent,openevolve}.

Recent related work on automated kernel generation and optimization also falls
into several categories. KernelEvolve, K-Search, AVO, and KernelFoundry
emphasize planning, autonomous variation, or quality-diversity search over
large kernel spaces \cite{kernelevolve,ksearch,avo,kernelfoundry}. CuTeGen,
TritonForge, PEAK, CudaForge, STARK, cuPilot, KernelBlaster, and KernelAgent instead structure optimization as iterative
refinement loops that combine generation with execution feedback, profiling,
hardware diagnosis, or persistent memory
\cite{cutegen,tritonforge,peak,cudaforge,stark,cupilot,kernelblaster,kernelagent,kernelagent_blog,deepseek_r1_blog}.
CUDAAgent, TritonRL, CUDA-LLM, KernelLLM, and cudaLLM move more capability
into specialized training or finetuning for CUDA or Triton generation
\cite{cuda_agent,tritonrl,cuda_llm,kernelllm,cudallm_model}, while
KernelFalcon, StitchCUDA, and Agentic Operator Generation broaden the scope to
correctness-first deep-agent generation, end-to-end GPU programs, or broad
operator coverage on new hardware \cite{kernelfalcon_blog,stitchcuda,agentic_asic}.
Compared with these systems, our focus is a resource-aware Triton optimizer
that promotes candidates by measured latency under multi-fidelity evaluation,
rather than relying primarily on offline RL, large evolutionary populations, or
backend-agnostic generation.

\section{Design}
\subsection{Problem Formulation}
Let $\mathcal{P}$ denote the space of candidate kernel programs reachable under
the allowed mutation operators, and let $x \in \mathcal{X}$ denote a concrete
tensor shape configuration (e.g., batch size, sequence length) drawn from
workload distribution $\mathcal{D}$.
For a candidate program $p \in \mathcal{P}$, define:
\begin{itemize}
  \item $C(p) \in \{0,1\}$: compile validity indicator,
  \item $K(p) \in \{0,1\}$: correctness indicator (passes task and numerical tests),
  \item $\ell(p; x)$: latency (us) on input $x$.
\end{itemize}

The target is to find the best valid program under correctness constraints:
\begin{equation}
  p^\star = \arg\min_{p \in \mathcal{P}} \; \mathbb{E}_{x \sim \mathcal{D}}[\ell(p;x)]
  \quad
  \text{s.t. } C(p)=1,\; K(p)=1.
\end{equation}

Equivalently, using throughput-style scoring used by the runtime loop, we can
maximize
\begin{equation}
  S(p,r) = \frac{10^6}{\max(1,\ell_r(p))} + \gamma(r),
  \label{eq:runtime-score}
\end{equation}
where $r$ is the \emph{rung} index, $b_r$ is the rung budget, and $\ell_r(p)$
is the measured kernel latency at fidelity level $r$. Inspired by similar concept of the work~\cite{asha}, 
a rung is a stage in a multi-fidelity evaluation ladder.
Successive rungs progressively increase measurement budget---more
warmup iterations, higher repeat counts, broader input shape coverage, and
optional GPU kernel profiling---and only top-performing candidates at each rung are
promoted to the next. The score $S(p,r)$ is dominated by latency: with latency
measured in microseconds, the leading term $10^6/\max(1,\ell_r(p))$
corresponds to executions per second. We can use
$\gamma(r)=\alpha r + \beta b_r$, where $\alpha$ and $\beta$ are small constants,
or any appropriate function $\gamma(r)$ that gives more weight to later rungs
to differentiate two candidates that have near-identical latency for scoring.  
$S(p,r)$ can be extended to include other factors 
such as code and optimization quality (e.g., judged by LLM), 
but we find that latency-based ranking is effective in practice.

Because only a finite set of candidates can be evaluated, search is budgeted.
Let $\mathcal{E}$ be the set of evaluated candidate-rung pairs and
$\mathrm{cost}(p,r)$ the measurement cost. The practical objective is:
\begin{equation}
  \max_{\mathcal{E}}\; \max_{(p,r)\in\mathcal{E}} S(p,r)
  \quad
  \text{s.t. } \sum_{(p,r)\in\mathcal{E}} \mathrm{cost}(p,r) \le B,
\end{equation}
with global budget $B$ (wall-clock and evaluation budget). This motivates
multi-fidelity scheduling: inexpensive low-rung screening followed by selective
promotion to high-fidelity evaluation.

\subsection{Definitions (Policy, Fidelity, and Runtime Terms)}
We use the following terms consistently throughout the paper.
\begin{itemize}
  \item \textbf{Policy check} $\mathcal{V}(p)$: a pre-benchmark gate that
  verifies the candidate Triton kernel compiles without error, exports the
  expected function signature, and satisfies output-shape and dtype constraints
  against the reference implementation. Only candidates that pass all checks
  proceed to benchmarking.
  \item \textbf{Fidelity level} $r$: the evaluation configuration used for a
  candidate, controlling the number of GPU warmup launches, timed repeat
  iterations, effort of GPU profiling, and the set of input shapes (e.g., batch size or
  sequence length) over which latency is measured.
  \item \textbf{Rung budget} $b_r$: the total measurement cost allocated to
  fidelity level $r$. Lower rungs use fewer repeats and a narrow shape
  subset for fast screening; higher rungs use more repeats, broader shape
  coverage, and optionally trigger GPU kernel profiling---using NVIDIA Nsight
  Compute (NCU), which reports per-kernel metrics such as SM occupancy, memory
  throughput, and pipeline stalls---for kernel analysis. Similar 
  profiling-driven diagnosis approach~\cite{kernelagent, ksearch} has been used to give LLM more useful hints.
  \item \textbf{Promotion decision}: the scheduler's determination of whether
  a candidate kernel at rung $r$ has sufficient performance to advance to
  rung $r+1$ for more thorough evaluation through ranking.
\end{itemize}

\subsection{Key Observations: Diversity-First Search with Progressive Evaluation}
\paragraph{Observation 1 (Search-space ruggedness).}
\label{sec:search-ruggedness}
Triton kernel optimization does not behave like smooth local search. A small
source-level edit---such as changing a tile size, reordering a reduction loop,
or adjusting a launch grid---can alter register pressure, shared-memory usage,
and the Triton compiler's code-generation decisions in non-obvious ways. Two
kernels that differ by a single line may exhibit vastly different compile
validity, numerical correctness, and runtime performance.

Purely evolutionary pipelines that repeatedly mutate a small elite set therefore
show high inter-round variance: a strong kernel at iteration $t$ can produce
descendants at $t+1$ that fail to compile or regress in latency. Over long
runs these failures compound---an incorrect candidate can pollute later mutation
context, a single noisy measurement can distort promotion decisions, and a
static mutation prior can trap exploration in a narrow region of the design
space. The practical implication is to prefer \emph{broad candidate generation
with targeted, expert-informed guidance} over narrow lineage evolution.

\paragraph{Observation 2 (Fidelity should follow quality).}
Not all kernel candidates deserve the same measurement effort. Beyond
low-level tuning, the choice of algorithm---e.g., online softmax vs.\
multi-pass reduction, or tiled vs.\ split-K matrix multiply---can
dominate performance differences, making it essential to explore algorithmic
alternatives early. A Triton kernel that crashes or produces wrong outputs can
be rejected in milliseconds, while confirming that a promising candidate is
genuinely faster requires many timed launches across diverse input shapes.
Allocating this expensive measurement uniformly wastes budget, so we couple
diverse, expert-guided candidate generation (across algorithm, tiling, compute/memory,
and launch choices) with a progressive multi-fidelity funnel: cheap early
rungs explore optimizations broadly while quickly prune invalid or weak kernels via correctness/compile gates,
while only strong survivors receive broader-shape, higher-repeat, optionally
profiled evaluation; this concentrates effort where it matters and improves
ranking robustness under timing noise.

%We therefore structure evaluation as a progressive funnel: early rungs use
%minimal warmup and few repeats to rapidly filter the candidate pool, and only
%top survivors advance to higher rungs with larger repeat counts, broader shape
%coverage, and optional GPU kernel profiling. Each promotion gate also enforces
%correctness and compilation checks, so invalid kernels are pruned before they
%consume measurement budget. Because noisy GPU timing can inflate or deflate a
%single measurement, concentrating repeats on fewer, already-vetted candidates
%produces more reliable rankings than spreading the same budget thinly across
%all candidates.

\subsection{System Architecture and Pipeline Overview}
Motivated by the observations above, our system is built around two principles:
(1) generate diverse candidates with expert-informed guidance with a focus on coverage---spanning
algorithmic alternatives, tiling strategies, memory layouts/accesses, computation decomposition, etc.---to cover the major optimization axes broadly that can help escape
brittle local descent, and (2) apply multi-fidelity scheduling not merely as a speed
optimization but also as a reliability mechanism---progressively stricter and more
expensive evaluation identifies kernels that are both fast and stable under
measurement noise.

\Cref{alg:pipeline} formalizes the optimization loop. We denote a candidate
by~$c$, rung index by~$r$, and the pending queue by~$Q$;
$\mathcal{V}(\cdot)$, $\mathcal{B}(\cdot)$, and $\mathcal{D}(\cdot)$
are policy verification, benchmark evaluation, and the scheduler's
promote-or-prune decision, respectively. Each iteration pops a
candidate--rung pair, verifies and benchmarks it, computes a latency-based
score (\cref{eq:runtime-score}), and queries $\mathcal{D}$ to decide
promotion or pruning. Promoted candidates re-enter the generation plane for
LLM code mutation and are re-enqueued at the next rung; pruned candidates are
discarded. The loop terminates when the queue empties, returning the best
kernel found.

\begin{algorithm}[t]
\caption{Online Multi-Fidelity Optimization Loop}
\label{alg:pipeline}
\begin{algorithmic}[1]
\Require Initial candidate queue $Q$, rung budgets $\{b_r\}$, fidelity policy $F(r)$
\State Initialize best candidate $c^\star \gets$ baseline, best score $s^\star \gets -\infty$
\While{$Q$ is not empty}
  \State Pop $(c, r, \text{strategy})$ from $Q$
  \State Update scheduler state for candidate $c$ at rung $r$
  \If{$\mathcal{V}(c)$ fails}
    \State Log rejection and \textbf{continue}
  \EndIf
  \State Evaluate $m \gets \mathcal{B}(c, r, F(r), b_r)$
  \If{$m$ failed}
    \State Log test failure and \textbf{continue}
  \EndIf
  \State Compute score $s(c,r)$ from adjudicated latency
  \State Profile and diagnose (GPU profiling/analysis/prompt context)
  \State Update scheduler with $(c,r,s(c,r))$
  \If{$s(c,r) > s^\star$}
    \State $c^\star \gets c$, $s^\star \gets s(c,r)$
  \EndIf
  \State $(\text{promote}, r^+) \gets \mathcal{D}(c,r,s(c,r))$ \Comment{$\mathcal{D}$ ranks and promotes candidates with scores}
  \If{promote}
    \State Construct next candidate code via the LLM mutation
    \State Push $(c, r^+, \text{strategy})$ to $Q$
  \EndIf
\EndWhile
\State \Return best candidate $c^\star$
\end{algorithmic}
\end{algorithm}

\clearpage

\begin{wrapfigure}{r}{0.5\textwidth}
\centering
\small
\resizebox{\linewidth}{!}{%
\begin{tikzpicture}[
  >=Latex,
  node distance=7mm and 8mm,
  block/.style={draw, rounded corners=2pt, align=center, fill=blue!4, minimum width=3.0cm, minimum height=7mm},
  decision/.style={draw, rounded corners=2pt, align=center, fill=green!6, minimum width=3.0cm, minimum height=7mm},
  io/.style={draw, rounded corners=2pt, align=center, fill=orange!8, minimum width=3.2cm, minimum height=7mm},
  data/.style={draw, rounded corners=2pt, align=center, fill=gray!12, minimum width=3.2cm, minimum height=7mm},
  flow/.style={-Latex, line width=0.6pt},
  feedback/.style={-Latex, line width=0.6pt}
]
  \node[io] (input) {Input Contract\\\texttt{test.py} + \texttt{input\_kernel.py}};
  \node[block, below=of input] (orch) {Orchestrator};

  \node[block, below left=11mm and 15mm of orch] (verify) {Verifier};
  \node[block, below=of verify] (bench) {Benchmarker};
  \node[block, below=of bench] (profile) {GPU Kernel Profiling +\\Diagnosis};

  \node[block, below right=11mm and 15mm of orch] (prompt) {Prompt Builder};
  \node[block, below=of prompt] (llm) {LLM Generation};
  \node[block, below=of llm] (repair) {Parser / Repair};

  \node[decision, below=19mm of orch] (sched) {Adaptive Resource Allocation\\(promote / prune)};
  \node[data, below=of sched] (store) {Run Storage +\\Checkpoints + Artifacts};
  \node[io, below=of store] (output) {Best Kernel Output\\\texttt{output\_kernel.py}};

  \draw[flow] (input) -- (orch);
  \draw[flow] (orch) -- (verify);
  \draw[flow] (verify) -- (bench);
  \draw[flow] (bench) -- (profile);

  \draw[flow] (orch) -- (prompt);
  \draw[flow] (prompt) -- (llm);
  \draw[flow] (llm) -- (repair);

  \draw[flow] (bench) -- (sched);
  \draw[flow] (profile) -- (sched);
  \draw[flow] (repair) -- (sched);
  \draw[flow] (sched) -- (store);
  \draw[flow] (store) -- (output);

  \draw[feedback] (sched.west) .. controls +(-1.1,0.2) and +(-0.6,-0.2) .. (verify.east);
  \draw[feedback] (sched.east) .. controls +(1.1,0.2) and +(0.6,-0.2) .. (prompt.west);

  \node[draw, rounded corners=2pt, inner sep=4pt, fit=(verify)(bench)(profile), label={[font=\scriptsize]above:Evaluation Plane}] {};
  \node[draw, rounded corners=2pt, inner sep=4pt, fit=(prompt)(llm)(repair), label={[font=\scriptsize]above:Generation Plane}] {};
\end{tikzpicture}%
}
\caption{System architecture of the kernel optimization agent.}
\label{fig:system-architecture}
\end{wrapfigure}
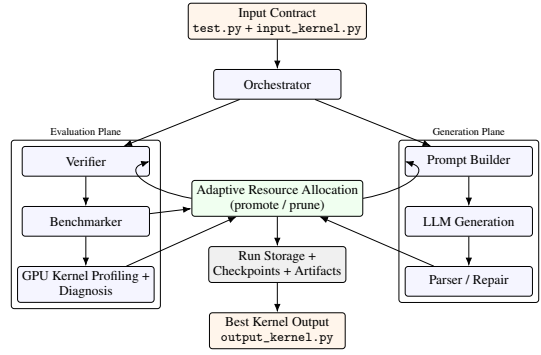
\Cref{fig:system-architecture} shows the end-to-end optimization and kernel evolution pipeline: the orchestrator
receives the input contract (\texttt{test.py} and \texttt{input\_kernel.py}),
routes each candidate through verification, benchmarking, and optional
profiling, then feeds diagnosis into the generation plane (prompt builder,
LLM, parser/repair). An adaptive resource allocation scheduler promotes or
prunes candidates at each rung while persisting all artifacts.
The architecture in \Cref{fig:system-architecture} is organized around two
parallel planes---an \emph{Evaluation Plane} and a \emph{Generation
Plane}---connected by a central adaptive resource allocation scheduler.

A run begins with two user-supplied files: a test file (correctness tests,
benchmark harness, input shapes) and an input kernel file (baseline Triton
kernel). The orchestrator seeds the candidate queue with the baseline and an
initial batch of expert-guided tasks.

\paragraph{Evaluation Plane.}
Each candidate flows through three stages. The \emph{Verifier} compiles the
kernel, validates its function signature, and checks output correctness against the
reference---failing candidates are rejected immediately. Survivors proceed to
the \emph{Benchmarker}, which measures GPU latency at the fidelity defined by
the current rung. At higher rungs, \emph{GPU Kernel Profiling} collects
hardware metrics via Nsight Compute and analyzes the bottleneck. 
These profiler signals serve two roles: (1) kernel diagnosis 
that identifies the efficiency level of the kernels from compute throughput and memory bandwidth perspectives,
and (2) prompt conditioning that steers the next mutation toward the
identified bottleneck and opportunities for improvement. 
%Importantly, ranking and diagnosis are intentionally
%separated---candidate promotion is driven by measured latency, while profiler
%evidence informs \emph{what} to mutate next.

Concretely, we track representative counters for each subsystem: compute and
memory saturation (SM/DRAM throughput compared to peak), execution mix
(FP32/FP16, tensorcore activity), occupancy (active warps), memory metrics
(global-load coalescing, cache hit rates, shared-memory bank conflicts), and
stall composition (barrier,
wait). Together, these counters provide a profile of current kernel's state of optimization.

\paragraph{Generation Plane.}
For promoted candidates, the \emph{Prompt Builder} assembles a structured
mutation prompt from the current kernel source, benchmark results, profiler
bottleneck analysis, prior round outcomes, and an expert strategy hint.
An example prompt construction is provided in Appendix~\ref{sec:appendix-prompt-template}.
This enables targeted edits (e.g., summary and proposed fixes based on profiling analysis) rather than unguided rewrites. Each round
follows a consistent schema---gather context, diagnose, build prompt, generate
code, parse/repair, and persist artifacts. The \emph{Parser / Repair} module extracts the kernel from the LLM
response and applies deterministic fixes (e.g., missing imports, broken
signatures) when the output is malformed; if parsing fails entirely,
deterministic fallback logic keeps the loop running.

At rung~0, the initial seeds--- used to
populate the initial candidate queue in Algorithm~1 and scheduled across many workers---can be summarized into
several broad mutation
families: \emph{memory-focused} edits (coalescing/reuse),
\emph{compute/pipeline-focused} edits (stage/overlap tuning), and
\emph{utilization-focused} edits (occupancy/parallel decomposition). This keeps
the first generation diverse but controlled, providing robust starting points
before richer diagnosis-conditioned mutations take over in later rounds.

\paragraph{Adaptive Resource Allocation and Search Control.}
Candidate generation draws from two sources:
expert-driven heuristic strategies for general GPU kernel optimization and LLM mutations conditioned on prior profiling diagnosis. The
scheduler ranks candidates by $s(c,r)$ with deterministic tie-breaking,
and online promotes the top fraction to the next rung with
increased budget, and prunes the rest. Promoted candidates re-enter the
evaluation plane at higher fidelity and also pass through the generation
plane to produce refined variants informed by richer profiling data. A key
design choice is the separation of ranking from mutation: promotion decisions
are budget-aware and scheduler-driven, while code edits are diagnosis-guided,
keeping the loop stable across long or interrupted runs.

\textbf{Key takeaway.} Conceptually, our agentic system operationalizes the workflow of an experienced GPU programmer: 
it begins with broad, low-cost exploration over diverse methods e.g., high-level algorithmic and implementation strategies, then progressively
 concentrates budget on the most promising candidates as evidence accumulates, with the assistance of profiling tools for analysis.

%\paragraph{Feedback and persistence.}
%Dashed arrows represent feedback: promoted candidates return to the verifier
%at higher fidelity, and profiling results flow to the prompt builder. The
%\emph{Run Storage} layer persists all artifacts---source code, logs, profiler
%reports, and prompt/response pairs. When the queue or budget is exhausted, the system
%emits the best kernel as \texttt{output\_kernel.py}.

% \subsection{Stability and Reproducibility}
% GPU timing measurements are inherently noisy, so the runtime includes an
% optional \emph{stability adjudication} step before each promotion: when a
% candidate's score change exceeds a configurable threshold, the system
% re-measures at the same fidelity level and reverts the candidate if the
% improvement is not confirmed. This serves two purposes---it protects ranking
% integrity against variance-induced false promotions and prevents unstable
% candidates from becoming mutation parents in later rounds. Every major
% action (generation, evaluation, promotion, diagnosis) is logged with full
% provenance for reproducible ablations and post-hoc analysis.

\section{Experiments}
\label{sec:eval}

We evaluate whether the system's design choices---expert-guided candidate
generation, multi-fidelity screening, and profiling-conditioned
mutation---translate into measurable gains on real Triton kernels. 
We evaluate six representative Triton benchmarks spanning LLM and traditional DNN workloads, including both standalone and fused kernels (Appendix~\ref{sec:appendix-benchmark-shapes}).
The experiments are organized around two questions: (1)~does the system produce similar or
faster kernels than baselines across workloads with different computational
profiles? and (2)~how does speedup evolve over the course of a
search run?

All experiments run on a single NVIDIA A100-80GB PCIe GPU with driver 535.261,
CUDA~12.8, PyTorch~2.9.1, Triton~3.5.1 on Ubuntu~20.04.
Across experiments, the LLM used for generation is GPT-5.4.
Each kernel is benchmarked in an isolated process with exclusive GPU access, implemented by the Evaluation plane. We
compare four methods: (1) the PyTorch reference (\texttt{torch}), (2) a
LLM-generated kernel without search (\texttt{base}) (See Appendix A.1), (3) an optimized kernel generated by
KernelAgent~\cite{kernelagent}, a state-of-the-art agentic kernel optimization system released by PyTorch, and (4) our system (\texttt{kernelbrain}).

\subsection{Candidate Variance}
Before presenting speedup comparisons, we first validate a core assumption
of our multi-fidelity design: that candidates generated by the same LLM exhibit large performance variance---the
rugged landscape described in Observation~1. This is also true for manual optimization: a kernel with careful optimization 
applied by an expert can be a lot faster than a na\"ive implementation.
(Section~\ref{sec:search-ruggedness}).

\begin{wrapfigure}{r}{0.48\textwidth}
\vspace{-1.0\baselineskip}
\centering
\includegraphics[width=0.94\linewidth]{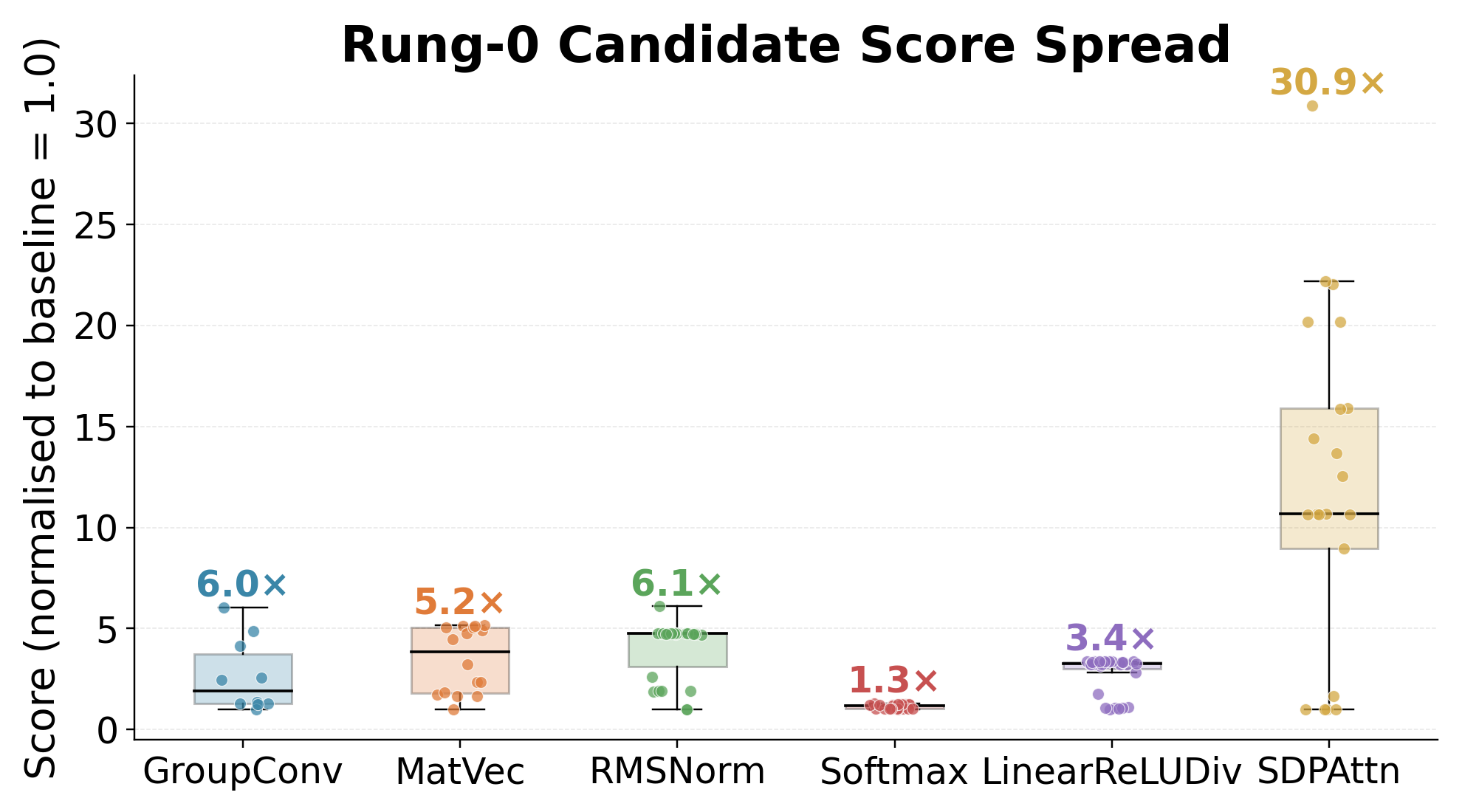}
\caption{Rung-0 candidate score spread across all six operators.
Scores are normalized to each benchmark's baseline ($=1.0\times$).
Each dot is one LLM-generated candidate evaluated. Box plots show quartiles; annotations report the
max-to-min ratio.}
\label{fig:score-spread}
\vspace{-1.3\baselineskip}
\end{wrapfigure}

\Cref{fig:score-spread} plots rung-0 per-candidate scores for all six
benchmarks, normalized to each benchmark's slowest. The spread is large
for several benchmarks: SDPAttn is widest ($30.9\times$), GroupConv and
RMSNorm remain broad ($>6\times$), and MatVec is also highly variable.
Softmax is comparatively tight ($\approx 1.3\times$), while LinearReLUDiv presents weak and strong clusters.
These distributions support our multi-fidelity design: cheap rung-0
screening has opportunity to remove very weak candidates early, and higher-fidelity budget is
reserved for the promising survivors.

\subsection{Speedup Results}
\label{sec:speedup}

%In these comparisons, \texttt{base} denotes the single-shot kernel from
%KernelAgent's \emph{generation} pipeline (KernelFalcon~\cite{kernelfalcon}), while \texttt{kernelagent} denotes
%KernelAgent's full \emph{optimization} pipeline. For MatVec and RMSNorm, we
%use the KernelAgent kernels released in their optimization examples.

\begin{figure*}[!htbp]
\centering
\includegraphics[width=\textwidth]{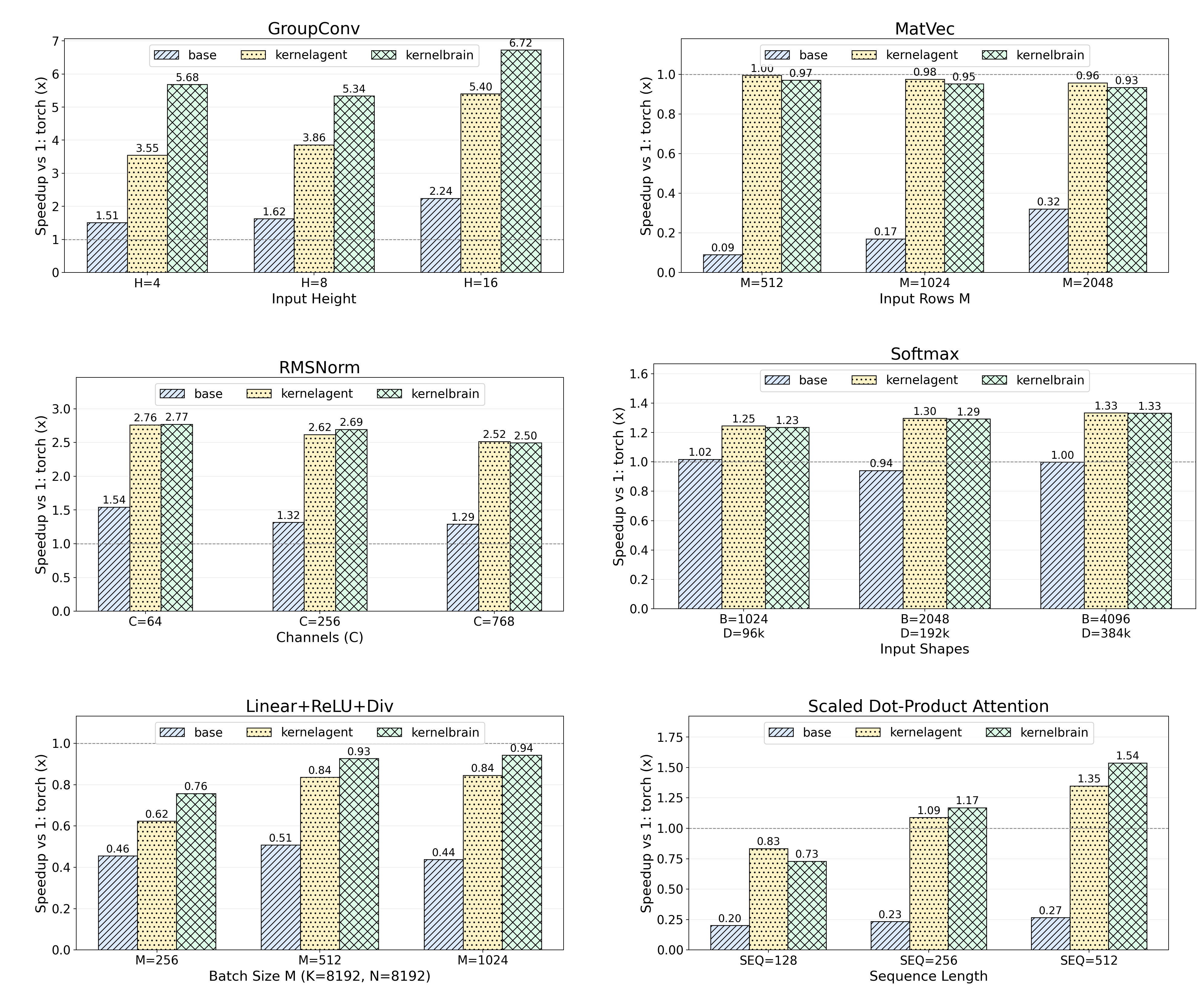}
\caption{Speedup comparison versus \texttt{torch} across all six benchmarks.
Row~1: (a)~GroupConv, a grouped 2-D convolution with 242 groups and kernel size $5{\times}768$, evaluated at input heights $H\in\{4,8,16\}$; and (b)~MatVec, a matrix--vector product $C{=}AB$ with $M\in\{512,1024,2048\}$ and fixed $K=256\text{k}$.
Row~2: (c)~RMSNorm, root-mean-square normalization over the channel dimension with $C\in\{64,256,768\}$; and (d)~Softmax, row-wise softmax with batch sizes up to 4096 and feature dimension up to $394\text{k}$.
Row~3: (e)~LinearReLUDiv, a fused linear layer followed by ReLU and element-wise division, with $K{=}N{=}8192$ and batch sizes $B\in\{256,512,1024\}$; and (f)~SDPAttn, scaled dot-product attention at sequence lengths $S\in\{128,256,512\}$ with HEAD\_DIM$=$1024.
Legend entries in each subplot are \texttt{base}, \texttt{kernelagent}, and \texttt{kernelbrain}.}
\label{fig:all-speedup}
\end{figure*}

\vspace{-0.45\baselineskip}

Across all six benchmarks, \texttt{kernelbrain} achieves the strongest overall
quality among compared methods, with the largest margins on GroupConv (which we used in one of production model deployment) and lower
optimization cost as detailed in Section~\ref{sec:search-efficiency}.

\Cref{fig:all-speedup} reports per-configuration speedups across all six
benchmarks. On GroupConv, \texttt{kernelbrain} achieves
$5.7\times$--$6.7\times$ over \texttt{torch} across input heights
$H\in\{4,8,16\}$, consistently outperforming both \texttt{base}
($1.5$--$2.2\times$) and \texttt{kernelagent} ($3.6$--$5.4\times$). The gap
widens at larger $H$, reaching $6.72\times$ at $H{=}16$, indicating that the
search discovers optimizations that scale with
problem size. 
%This behavior is consistent with the coarse-to-fine strategy:
%early rounds seed diverse algorithmic options (e.g., tiling, persistent
%kernels), and later rounds refine them using profiling feedback. 
On MatVec, all methods perform near the \texttt{torch} baseline
($0.93$--$1.0\times$).
For the remaining benchmarks, the same pattern holds with workload-dependent
headroom: on RMSNorm, \texttt{kernelbrain} reaches $2.5$--$2.8\times$ and
matches \texttt{kernelagent} while clearly outperforming \texttt{base}
($1.3$--$1.5\times$); on Softmax, both search-based methods cluster at
$1.25$--$1.33\times$ and \texttt{base} stays on par; on LinearReLUDiv,
\texttt{kernelbrain} improves to $0.76$--$0.94\times$ versus
\texttt{kernelagent} ($0.62$--$0.84\times$) and \texttt{base}
($0.46$--$0.44\times$), reducing but not eliminating the gap to PyTorch; and
on SDPAttn, \texttt{kernelbrain} leads at larger sequence lengths with
$1.17$--$1.54\times$ versus \texttt{kernelagent} ($1.09$--$1.35\times$),
while both methods remain below \texttt{torch} at $S{=}128$ due to launch
overheads. Though some bases are very slow (e.g., SDPAttn and MatVec), 
the eventual results demonstrate the effectiveness of our approach to evolve kernels towards faster implementations. 
Overall, \Cref{fig:all-speedup} shows that multi-fidelity, profiling-guided
search consistently improves over \texttt{base} and usually over
\texttt{kernelagent}, with gains on benchmarks that expose richer
algorithmic and scheduling choices.

\subsection{Common Optimization Patterns Across Benchmarks}
Although the six benchmarks differ substantially in arithmetic intensity and
dataflow, the best kernels converge to a small set of recurring GPU
optimization patterns. First, the largest gains come from \emph{changing the
parallel decomposition of work}, not from minor local edits. For example,
MatVec moves from a serial inner-product reduction to a cooperative split
reduction across threadblocks, while SDPAttn improves by partitioning the head
dimension into smaller slices that lower the live working set and increase
threadblock residency. These changes expose more useful parallel work to the
SMs and reduce the chance that a single oversized program becomes limited by
register pressure or poor occupancy.

Second, several winners improve primarily by \emph{reducing global
memory traffic}. Softmax replaces an always-two-pass implementation with a
shape-aware fused or online reduction, and RMSNorm introduces a fast path that
keeps small normalization tiles on chip rather than rereading them from memory.
In both cases, the algorithmic objective is unchanged, but the optimized dataflow
better matches the GPU memory hierarchy by converting redundant memory passes
into buffered or streaming reductions.

Third, for fused dense kernels such as LinearReLUDiv and for kernels
such as GroupConv, the dominant improvements come from \emph{better scheduling
and locality management}. The strongest variants use more appropriate launch
topologies, tile shapes, and code paths, which improve cache reuse,
reduce boundary overhead, and align the program layout more closely with the
underlying workload geometry. 

Appendix~\ref{sec:appendix-softmax-shortlist} shows some sample LLM generated code change recommendations
during an optimization run.
%Overall, these results suggest that expert-guided
%#search is most valuable when it can traverse a space of high-level GPU design
%choices---work partitioning, reduction structure, memory traffic shaping, and
%launch organization---rather than only tuning scalar hyperparameters around a
%fixed kernel skeleton. 

\textbf{Key takeaway.} These observations from the generated kernels again suggest that agentic GPU kernel optimization 
is most effective when it explores broad transformations that alter high-level kernel structure, 
including work partitioning, reduction organization, memory-traffic scheduling, and launch configuration.  
By exploring more GPU optimization techniques, this opens the door to larger performance gains than would be possible by spending budget heavily iteratively on small number of workers which tends to stagnate easily~\cite{openevolve, ksearch}.
The \texttt{kernelagent} uses small number of workers and more iteration counts. But the work tried to solve the diversity problem using \emph{Beam Search} at each step. 
As shown in the discussed experiments results (\Cref{sec:speedup}), \texttt{kernelbrain} still is able to find better kernels for some benchmarks, while consuming much less overall search time (See Section \Cref{sec:search-efficiency}).

\subsection{Observed Run Trajectory}
\begin{wrapfigure}[16]{r}{0.44\textwidth}
\centering
\scriptsize
\resizebox{0.95\linewidth}{!}{%
\begin{tikzpicture}[
  >=Latex,
  node distance=5mm and 7mm,
  stage/.style={draw, rounded corners=2pt, align=center, fill=blue!4, minimum width=4.1cm, minimum height=6mm},
  summary/.style={draw, rounded corners=2pt, align=center, fill=green!6, minimum width=4.1cm, minimum height=6mm},
  flow/.style={-Latex, line width=0.6pt}
]
  \node[stage] (start) {Run Start};
  \node[stage, below=of start] (gen) {Seed + Generate\\20 candidates + baseline};
  \node[summary, below left=of gen] (r0) {Rung 0 eval, budget=small\\Success: 16\\Fail: 5};
  \node[summary, right=of r0] (p1) {Promoted to Rung 1: 7 total\\(6 candidates + baseline)};
  \node[summary, below=of r0] (r1) {Rung 1 eval, budget=medium\\Success: 5\\Fail: 2};
  \node[summary, right=of r1] (p2) {Promoted to Rung 2: 3 total\\(2 candidates + baseline)};
  \node[summary, below=of r1] (r2) {Rung 2 eval, budget=large\\Success: 3\\Fail: 0};
  \node[stage, right=of r2] (done) {Run Finish\\Best: cand\_0013\\score 384.36\\speedup 5.30x};

  \draw[flow] (start) -- (gen);
  \draw[flow] (gen.south west) -- (r0.north);
  \draw[flow] (r0) -- (p1);
  \draw[flow] (p1.south west) -- (r1.north east);
  \draw[flow] (r1) -- (p2);
  \draw[flow] (p2.south west) -- (r2.north east);
  \draw[flow] (r2) -- (done);

\end{tikzpicture}%
}
\caption{Observed run trajectory with explicit promotion counts and representative candidate outcomes.}
\label{fig:run-trajectory}
\end{wrapfigure}
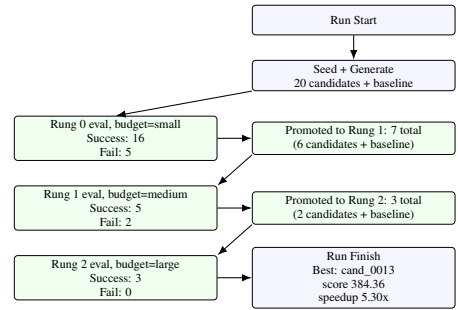

\vspace{0.2\baselineskip}

\Cref{fig:run-trajectory} illustrates one complete sample optimization run (MatVec)
to show how the multi-fidelity funnel allocates budget. Starting from 20
candidates plus baseline, rung~0 quickly prunes failures (16 pass, 5 fail),
and only 7 candidates are promoted to rung~1. After stricter re-evaluation,
2 more are dropped and 3 survivors reach rung~2 (full budget with profiling);
all 3 pass, and the best reaches $5.30\times$ over baseline. Detailed per-rung survival statistics and interpretation is shown in
Appendix~\ref{sec:appendix-rung-breakdown} (\Cref{tab:rung-breakdown}).

Most failed jobs shown in the table arise from more challenging source-level edits.
In simple failure cases, explicit policy guards improve generation quality.
For example, enforcing test-harness constraints, preserving the required function
signature, staying within Triton/Python rather than switching languages, and
avoiding unsupported third-party calls, optionally with short code exemplars,
helps steer the LLM toward valid candidates.

\begin{figure}[!t]
\centering
\includegraphics[width=0.88\linewidth]{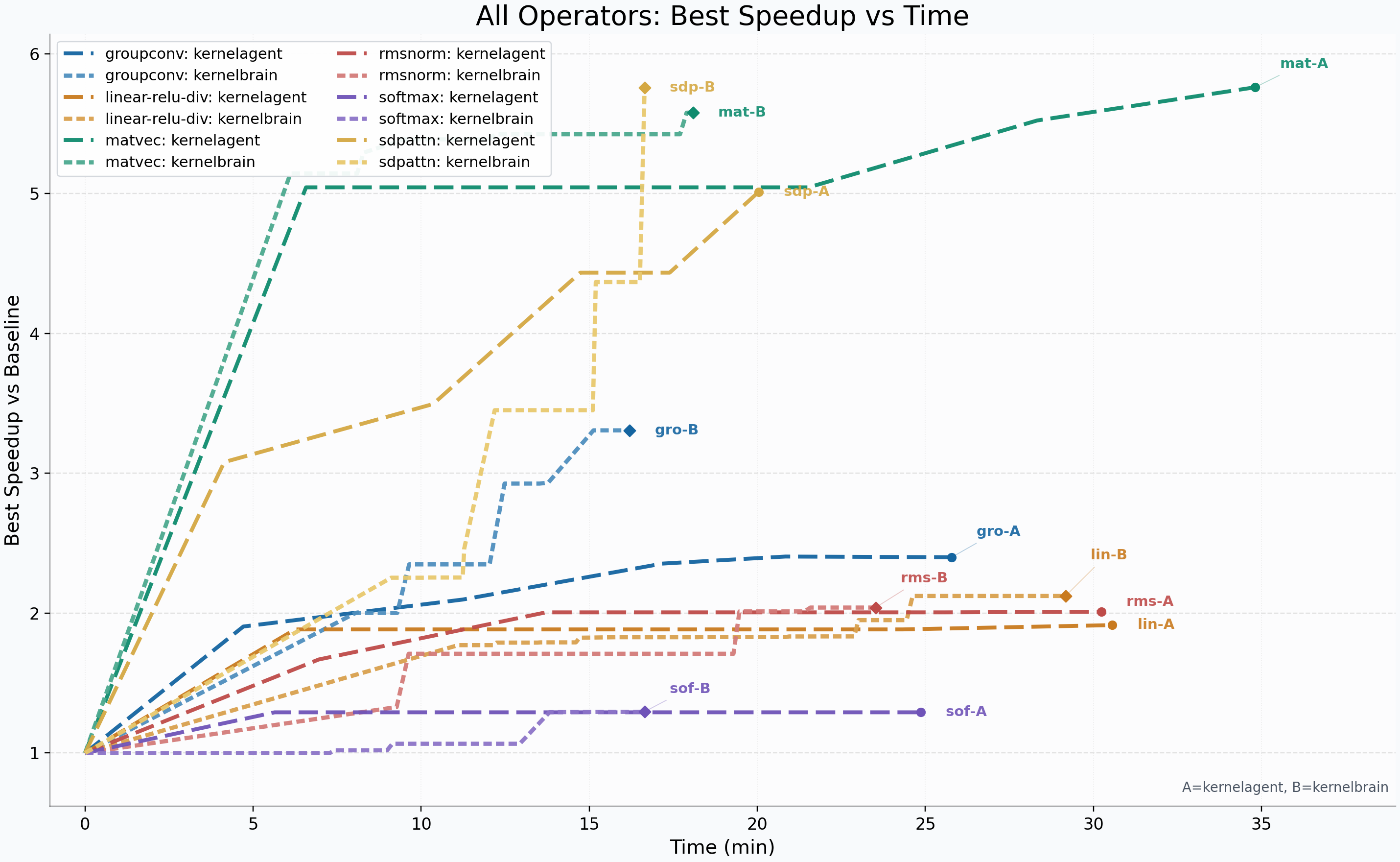}
\caption{Combined best-speedup-versus-time trajectories across all six benchmarks. Each benchmark uses one color family; dashed styles distinguish \texttt{kernelagent} and \texttt{kernelbrain}. The x-axis reports end-to-end
optimization runtime in minutes}
\label{fig:all-operators-perf-time}
\end{figure}

\vspace{-0.35\baselineskip}

\subsection{Search Efficiency over Time}
\label{sec:search-efficiency}
\Cref{fig:all-operators-perf-time} overlays the best-speedup-over-time
trajectories for all six benchmarks starting from \texttt{base}. Though fair
comparison across different agent architectures remains an open research
question, we report a controlled wall-clock comparison under a common budget
setting. For each benchmark, we set the global budget $B$ (in Eq.~(3)) of the
full optimization run to up to 30 minutes of search time. \texttt{Kernelagent}
does not allocate uniform budget across runs with \emph{Beam Search}.
GroupConv show rapid early
gains while
MatVec converge quickly to better performance.
LinearReLUDiv follows a steady upward curve, with \texttt{kernelbrain}
reaching its best speedup ($2.14\times$) in roughly the same wall-clock time
that \texttt{kernelagent} plateaus at $1.90\times$.
SDPAttn exhibits a distinctive step-function trajectory: sharp jumps occur after which speedup climbs steadily
to $5.7\times$.
Together with the speedup results
above, these trajectories demonstrate that the combination of expert-guided
generation and multi-fidelity scheduling used in \texttt{kernelbrain} produces strong kernels effectively and efficiently.

\begin{wrapfigure}[11]{r}{0.44\textwidth}
\vspace{-0.45\baselineskip}
\centering
\includegraphics[width=0.92\linewidth]{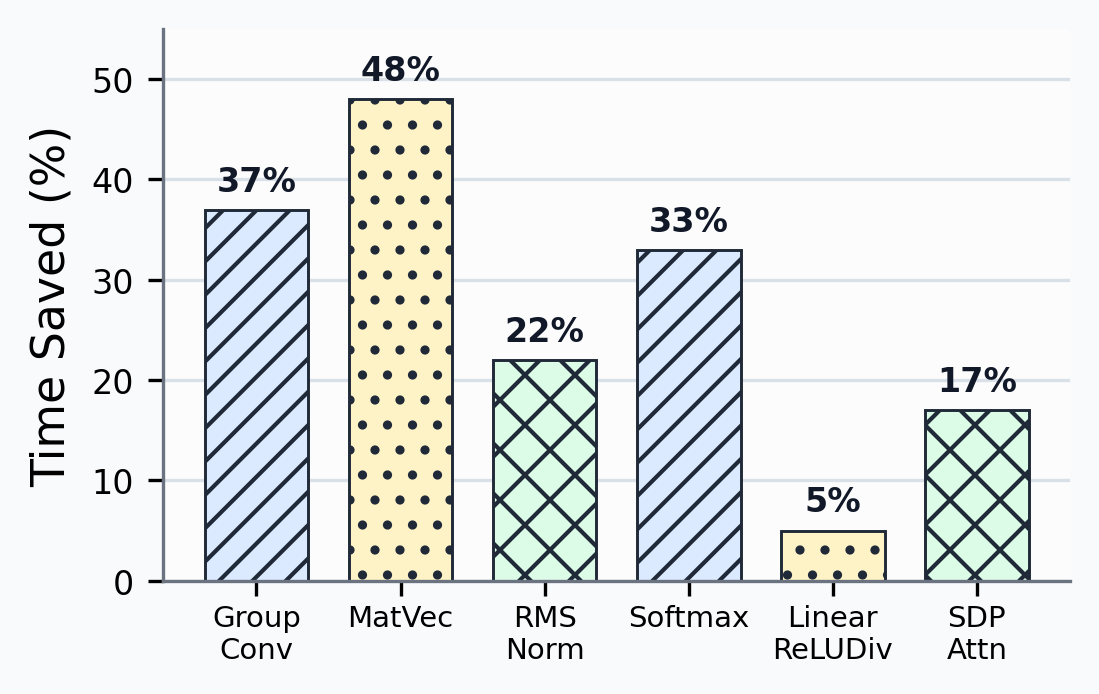}
\captionsetup{hypcap=false}
\caption{KernelBrain (KB) time saved versus KernelAgent (KA);}
\label{fig:kb-vs-ka-time-saved}
\end{wrapfigure}

\Cref{fig:kb-vs-ka-time-saved} quantifies the optimization-time advantage,
Across all six benchmarks, \texttt{kernelbrain} matches or
exceeds \texttt{kernelagent} kernel quality, while consuming
significantly less wall-clock time---saving 17--48\% on five of six
operators. Even on LinearReLUDiv, where the time saving is modest (5\%),
\texttt{kernelbrain} still produces a 12\% faster kernel.
Together with the trajectories in \Cref{fig:all-operators-perf-time}, these
results confirm that multi-fidelity screening and expert-guided mutation
yield both better kernels and shorter optimization runs.

% \begin{table}[t]
% \centering
% \caption{Example result table.}
% \begin{tabular}{lccc}
% \toprule
% Method & Mean (ms) & Std (ms) & Speedup \\
% \midrule
% Baseline & 7.72 & 0.01 & 1.00$\times$ \\
% Ours & 4.66 & 0.01 & 1.66$\times$ \\
% \bottomrule
% \end{tabular}
% \end{table}

\section{Conclusion}

We presented KernelBrain, a practical agentic optimizer for Triton kernel that
combines expert-guided candidate generation, adaptive multi-fidelity resource
allocation, and profiling-conditioned mutation under strict correctness and
measurement controls. This integration improves both kernel quality and search efficiency, supporting our central
claim that robust kernel optimization depends on jointly designing scheduling, diagnosis, and resource allocation.

\textbf{Limitations.} Our current optimization scope is source-level Triton kernel generation and
editing. Because Triton backend compilation and low-level code generation are
outside this system's control, some performance opportunities may remain
unexploited even when source-level hints are strong. In practice, this means
the framework improves kernels primarily through better algorithmic/scheduling
guidance at the source level, while backend-level improvements require future
co-design with compiler/runtime components.

\textbf{Broader Impact.} This work can have positive impact by reducing the cost and energy required to
optimize high-performance kernels, which can improve accessibility of efficient
ML systems. It can also reduce optimization turnaround time in production ML engineering and lower
infrastructure cost for repeated kernel tuning and deployment. The idea is also generic and could be applied to other agentic domains.

\clearpage
\appendix
\setcounter{table}{0}
\renewcommand{\thetable}{A.\arabic{table}}
\section{Appendix}

\subsection{Benchmark Descriptions and Shapes}
\label{sec:appendix-benchmark-shapes}
Below is a detailed description of the benchmarks evaluated in this paper.

\noindent\textbf{GroupConv.} Grouped 2-D convolution matching
\texttt{nn.Conv2d} with groups$=242$, in\_channels$=242$,
out\_channels$=217{,}800$, kernel size $5{\times}768$, stride $(1,1)$,
padding $(4,0)$, and dilation $(1,1)$. The benchmark sweeps input tensors of
shape $(N,C,H,W)=(1,242,H,768)$ with $H\in\{4,8,16\}$.

\noindent\textbf{MatVec.} Matrix--vector multiplication $C{=}AB$ with
$A\in\mathbb{R}^{M\times K}$ and $B\in\mathbb{R}^{K\times 1}$. The benchmark
sweeps $M\in\{512,1024,2048\}$ with fixed $K=256\text{k}\;(=2^{18})$.

\noindent\textbf{RMSNorm.} Root-mean-square normalization over the channel
dimension of an NCHW tensor. The benchmark sweeps inputs of shape
$(N,C,H,W)=(16,64,S,S)$ with $S\in\{64,256,768\}$ and uses
$\varepsilon=10^{-5}$.

\noindent\textbf{Softmax.} Row-wise softmax over the feature dimension. The
benchmark uses three labeled points:
$(1024,96\text{k}\;(=3\cdot 2^{15}))$, $(2048,192\text{k}\;(=3\cdot 2^{16}))$, and $(4096,384\text{k}\;(=3\cdot 2^{17}))$ for
$(\text{batch\_size},\text{dim})$.

\noindent\textbf{LinearReLUDiv.} Fused linear transform, ReLU, and divide by a
constant. The benchmark sweeps input matrices $x\in\mathbb{R}^{M\times 8192}$
with $M\in\{256,512,1024\}$, fixed $K=N=8192$, and divisor $2.0$.

\noindent\textbf{SDPAttn.} Scaled dot-product attention with
$Q,K,V\in\mathbb{R}^{32\times 32\times S\times 1024}$. The benchmark sweeps
sequence length $S\in\{128,256,512\}$ with batch size $32$, $32$ heads, and
head dimension $1024$.

For GroupConv, Softmax, LinearReLUDiv, and SDPAttn, the
base kernels are generated by the KernelFalcon kernel-generation pipeline
\cite{kernelfalcon_blog}. For MatVec and RMSNorm, we use kernels from
KernelAgent's released examples as base kernels \cite{kernelagent}.

\subsection{Per-Rung Survival Statistics}
\label{sec:appendix-rung-breakdown}
\Cref{tab:rung-breakdown} summarizes per-rung candidate survival across all
six benchmarks. In our paper runs, we typically found a 3-rung schedule
sufficient, with 20--30 initial candidates (not including the base). The runtime score uses
$\alpha=0.02$ and $\beta=0.01$, and the measurement iterations increase from
R0 to R2 as $4 \rightarrow 16 \rightarrow 60$.
As shown in \Cref{tab:rung-breakdown}, only 19.0--35.5\% advance to rung~1,
with the remainder filtered by test failures and ranking at Rung~0. A second
round of pruning at rung~1 leaves 3--5 finalists---roughly 9.7--19.0\% of the
original pool---who receive full-budget rung-2 evaluation. The overall
Rung~0 test-failure rate ranges from 0\% (LinearReLUDiv and Softmax) to
52.4\% (GroupConv), reflecting workload-dependent kernel-generation
difficulty.

\begin{table}[H]
\centering
\caption{Per-rung candidate survival across the six benchmarks.
Fail\% is relative to the number of candidates entering that rung;
Promotion\% is relative to the original candidate count.}
\label{tab:rung-breakdown}
\footnotesize
\begin{tabular}{lcccc}
\toprule
\textbf{Benchmark} & \textbf{R0 Fail\%} & \textbf{R0$\to$R1\%} & \textbf{R1 Fail\%} & \textbf{R1$\to$R2\%} \\
\midrule
GroupConv        & 52.4\% & 19.0\% & 25.0\% & 14.3\% \\
MatVec           & 23.8\% & 33.3\% & 28.6\% & 14.3\% \\
RMSNorm          & 16.1\% & 25.8\% & 25.0\% &  9.7\% \\
Softmax          &  0.0\% & 19.0\% &  0.0\% & 19.0\% \\
LinearReLUDiv    &  0.0\% & 35.5\% &  9.1\% & 16.1\% \\
SDPAttn          & 19.2\% & 23.1\% &  0.0\% & 15.4\% \\
\bottomrule
\end{tabular}
\end{table}

\subsection{Optimization Prompt Template}
\label{sec:appendix-prompt-template}
The optimization prompt is assembled from the
baseline analysis, optimization prescription, benchmark summary, policy
constraints, and baseline kernel source; later rungs additionally populate a
combined profiling summary, proposed fixes etc. The LLM can use the
profiling results to suggest prescriptions for further improvement.

\newpage
\begin{center}
\textbf{Abridged prompt skeleton.}
\end{center}
\vspace{-0.5em}
\begin{minipage}{\linewidth}
\begin{lstlisting}[style=promptstyle]
You are a Triton kernel optimization expert.

===========================
1) Objective
===========================
Produce a better kernel.py candidate while preserving correctness...

===========================
2) Candidate Metadata
===========================
candidate_id: {candidate_id}

===========================
3) Input Context
===========================
Baseline Analysis (JSON):
{baseline_json}

Optimization Prescription:
  category:        {prescription_category}
  summary:         {prescription_summary}
  assessment:      {prescription_assessment}
  proposed_change: {prescription_proposed_change}

Kernel Synthesis Target:
{synthesis_target}

Policy Constraints:
{policy_constraints}

Recent Benchmark Metrics (JSON):
{benchmark_json}

Recent Profiling Summary (JSON, NCU + profiling analysis):
  ncu counters:      {ncu_json}
  profiling_analysis: {profiling_json}

Recent Debug Logs:
{debug_logs}

Current Kernel Source (Python):
{kernel_code}

===========================
4) Output Requirements
===========================
- Return only full Python source code for the mutated kernel module.
- Preserve the callable kernel_function entrypoint and compatibility.
- Apply at least one substantive performance-oriented code change.
...
\end{lstlisting}
\end{minipage}

\subsection{Softmax Optimization Prompt Shortlist}
\label{sec:appendix-softmax-shortlist}
\Cref{tab:softmax-prompt-shortlist} summarizes the observed optimization fixes
proposed by the AI agent for Softmax during the search process. The prompts are
grouped to cover complementary optimization categories (kernel formulation,
parallel decomposition, layout/stride strategy, tiling/on-chip residency, and
autotuning/occupancy).

\footnotesize
\setlength{\tabcolsep}{4pt}
\begin{longtable}{p{0.06\linewidth}p{0.89\linewidth}}
\caption{AI-observed Softmax proposed fixes across optimization categories.}
\label{tab:softmax-prompt-shortlist}\\
\toprule
\textbf{ID} & \textbf{Prompt} \\
\midrule
\endfirsthead

\toprule
\textbf{ID} & \textbf{Prompt} \\
\midrule
\endhead

\bottomrule
\endfoot

\multicolumn{2}{l}{\textbf{Kernel Formulation and Math}} \\
1 & Replace the two-kernel design with a fused row-wise softmax kernel using online normalization. \\
2 & Use an online/single-pass softmax reduction algorithm for row max and normalization stats. \\
3 & Use exp2-based formulation with scaling where numerically acceptable. \\
4 & Replace division by row\_sum with multiplication by reciprocal when profitable. \\
5 & Keep reductions in fp32 but consider narrower IO and vectorized loads/stores for contiguous cases. \\

\addlinespace[2pt]
\multicolumn{2}{l}{\textbf{Parallel Decomposition and Mapping}} \\
6 & Use shape-dependent parallelization: multiple rows per program for small rows, multiple programs per row for very large rows. \\
7 & Use a larger grid decomposition for long rows, e.g., split one row across multiple programs/warps and reduce partials. \\
8 & Use a multi-program or hierarchical reduction strategy for very large n\_cols or low n\_rows workloads. \\
9 & Introduce a 2D or split-row parallelization strategy for large-column softmax. \\
10 & Batch multiple small rows per program when n\_cols is small. \\
11 & Use persistent row-level scheduling for contiguous inputs. \\

\addlinespace[2pt]
\multicolumn{2}{l}{\textbf{Layout and Stride Strategy}} \\
12 & Ensure the softmax dimension is contiguous or add optimized kernels for common strided layouts. \\
13 & Prefer making the reduced dimension contiguous before launching softmax when tensor reuse or batch size makes the transform worthwhile. \\

\addlinespace[2pt]
\multicolumn{2}{l}{\textbf{Tiling and On-Chip Residency}} \\
14 & Add a specialized path for small/medium n\_cols where an entire row can stay in SRAM/registers. \\
15 & If n\_cols fits, compute softmax in a single tiled resident pass with on-chip buffering. \\
16 & For large rows, tile through shared/SRAM and compute partial results to avoid full-row rereads when feasible. \\
17 & Introduce a specialized one-block softmax kernel for rows that fit entirely in a tile. \\
18 & Increase coverage of the single-pass contiguous-row kernel beyond the current <= 2048 threshold where hardware resources allow. \\

\addlinespace[2pt]
\multicolumn{2}{l}{\textbf{Autotuning, Occupancy, and Validation}} \\
19 & Add Triton autotuning for BLOCK\_SIZE, num\_warps, and num\_stages keyed on n\_cols and dtype. \\
20 & Autotune smaller BLOCK\_SIZE options such as 128/256/512 for large-row cases. \\
21 & Retune/autotune with occupancy-aware constraints, including smaller BLOCK\_SIZE or lower register-pressure variants. \\
22 & Profile register usage and occupancy with NCU/Triton compiler reports, then prune configs that depress residency. \\
23 & Add shape-specific heuristics to cap BLOCK\_SIZE near the smallest power-of-two covering n\_cols instead of allowing overly large tiles. \\
\end{longtable}
\normalsize

\subsection{Relative Variability Metrics}
\label{sec:appendix-relative-variability}
\Cref{tab:kernelbrain-tail-p50-run1} reports KernelBrain kernel execution time variability as
normalized percentile ratios (p90/p50, p95/p50, p99/p50) for all benchmark
shapes, when measured over the iterations of each benchmark run. Across these
rows, p90/p50 ranges from 1.002240 to 1.076765, p95/p50 ranges from 1.003646
to 1.135351, and p99/p50 ranges from 1.008960 to 1.253867 (25\% diff). The main point is
that kernel ranking itself becomes challenging under this variability:
benchmark- and shape-dependent tail behavior can materially reorder close
candidates, so variance must be treated as a ranking-stability issue rather
than conflated with speedup magnitude.For each shape, the speedup is computed as below,
$\bar{t}_{\text{shape}}=\frac{\sum_{i=1}^{N_{\text{iter}}} t_i}{N_{\text{iter}}}$,
and,
$\mathrm{speedup}_{\text{shape}}=\bar{t}_{\text{base,shape}}/\bar{t}_{\text{opt,shape}}$.

\begin{table}[H]
\centering
\scriptsize
\setlength{\tabcolsep}{4pt}
\caption{KernelBrain tail ratios normalized by p50.}
\label{tab:kernelbrain-tail-p50-run1}
\begin{tabular}{llccc}
\toprule
Benchmark & Shape & p90/p50 & p95/p50 & p99/p50 \\
\midrule
Softmax & (1024, 96k) & 1.014571 & 1.018561 & 1.031527 \\
Softmax & (2048, 192k) & 1.005321 & 1.007390 & 1.008960 \\
Softmax & (4096, 384k) & 1.002240 & 1.003646 & 1.010635 \\
LinearReLUDiv & M=256 & 1.030170 & 1.040696 & 1.049231 \\
LinearReLUDiv & M=512 & 1.023364 & 1.027099 & 1.048125 \\
LinearReLUDiv & M=1024 & 1.033003 & 1.040451 & 1.048637 \\
RMSNorm & S=64 & 1.076765 & 1.135351 & 1.253867 \\
RMSNorm & S=256 & 1.053016 & 1.074292 & 1.093731 \\
RMSNorm & S=768 & 1.040643 & 1.051436 & 1.070240 \\
GroupConv & H=4 & 1.011511 & 1.013012 & 1.025040 \\
GroupConv & H=8 & 1.011198 & 1.019219 & 1.021929 \\
GroupConv & H=16 & 1.023299 & 1.032570 & 1.033772 \\
ScaledDotProduct & SEQ=128 & 1.011675 & 1.016662 & 1.180335 \\
ScaledDotProduct & SEQ=256 & 1.011468 & 1.013225 & 1.023878 \\
ScaledDotProduct & SEQ=512 & 1.009772 & 1.011608 & 1.014925 \\
MatVec & M=512 & 1.073976 & 1.090014 & 1.112247 \\
MatVec & M=1024 & 1.026827 & 1.030629 & 1.051708 \\
MatVec & M=2048 & 1.015673 & 1.021878 & 1.028976 \\
\bottomrule
\end{tabular}
\end{table}

\clearpage

\end{document}